**Title:** How do causal assumptions work? A step-by-step guide

**Standfirst:** Causal assumptions are often used in health sciences but how do they allow us to estimate causal effects from observed data? Using a hypothetical example, we follow, step-by-step, how causal assumptions can bring us from an observed estimate to a causal one. This tutorial is targeted at readers who want to know what assumptions they rely on when estimating causal effects using confounder control and how those assumptions, when satisfied, allow us to interpret estimates as causal.

**Author:** Labrecque JA[1]

[1]Department of Epidemiology, Erasmus MC, Rotterdam, the Netherlands

**ORCID:** 0000-0002-1028-2158

**Summary points:**

- Causal inference works by clarifying what a causal question is and how, via assumptions, we can connect a causal question to observed data.
- Estimating causal effects by confounder adjustment requires four causal assumptions: consistency, exchangeability, positivity and no measurement error
- Each assumption plays a different role in connecting causal quantities to observed data.
- This article takes the reader step-by-step from observed data to causal estimate.

## Introduction

Causal inference is about formulating clear causal questions, using appropriate estimators to answer those questions and recognizing which causal assumptions are required to allow causation to be inferred. The causal assumptions specifically play an important role in clarifying the conditions under which an estimate obtained can be interpreted as causal. In health sciences, the most common set of causal assumptions is that used to estimate causal effect by controlling confounding, namely, consistency, conditional exchangeability, positivity and no measurement error. Many introductory articles and textbooks have been written defining and explaining these assumption[1,2]. None, however, provide the reader step-by-step intuitions for how these assumptions perform this almost magical task of obtaining causal conclusions from observable data. Here, using a hypothetical example where we want to estimate the effect of a fictitious prenatal program on neonatal mortality, we demonstrate how each assumption is one step from observable data to a causal estimate.

This article is intended for readers new to causal inference. Though, it contains mathematical expressions, they are all explained in words and care has been taken so that the intuition this article intends to provide is accessible without them. It should still be mentioned that many of the concepts discussed are not the type that sink in immediately. Therefore, we advise the reader not to worry if they do not immediately grasp a concept.

*What are causal effects?*

Before we get into causal assumptions, we must be precise about what we mean by a causal effect. Imagine we had a prenatal program, which we will call Program A, and we want to know whether it helped reduce neonatal mortality. We might answer this question by finding a group of hospitals which used Program A and compare the neonatal mortality in those hospitals to the neonatal mortality in hospitals that did not use Program A. If we find that neonatal mortality was lower in Program A hospitals, does that mean Program A reduces neonatal mortality? No. Regardless of how similar the hospitals in each group might be, we do not know what would have happened to Program A hospitals had they not used Program A or what would have happened to non-Program A hospitals had they used Program A.

Take a second to reread the last sentence of the previous paragraph. Note that it does not compare Program A and non-Program A hospitals with each other. It compares each group of hospitals with what would have happened to that same group of hospitals at the same time but in the condition we did not observe. This is how we define and think about causal effects: by comparing what would have happened to the same group of hospitals (or people as the case may be) under different, often unobserved, scenarios. These scenarios of what could have been observed had a different action been taken are called potential outcomes [3]. They are outcomes we could have potentially observed. If we knew the two potential outcomes for all hospitals in our case, the neonatal mortality had all hospitals used Program A and the neonatal mortality had all hospitals not used Program A, we would know if Program A had a causal effect because if it did, the potential outcomes would be different.

You may be thinking, if estimating causal effects require comparing the same group of hospitals or people in two conditions, why then do most epidemiologic studies, even randomized controlled trials, compare two *different* groups of people, one treated and one untreated? Quite simply, we do this because it is the only thing we can do. How do we go, then, from comparing two different groups of people or hospitals in two different conditions to saying something about potential outcomes which we have just admitted are impossible to observe?

*Causal inference to the rescue*

Causal inference bridges this gap between observable data and potential outcomes. Let us begin with what we can observe. We can measure the neonatal mortality in Program A hospitals and the neonatal mortality in non-Program A hospitals and get a risk difference:

$$\Pr(Y|A=1)-\Pr(Y|A=0) \tag{1}$$

Where $\Pr(Y|A=1)$ is the proportion of births resulting in neonatal mortality (which we will label $Y$) among hospitals who use Program A (which we will label A=1). $\Pr(Y|A=0)$ is the same proportion among non-Program A hospitals. The vertical bar '|' can be thought of as "among". When you see '|' you know we are looking in a subgroup of hospitals. The risk difference above, therefore, compares two different groups of hospitals: those that used Program A and those that did not.

This is, as we have already established, is not what we want to know. What we want to know, the causal effect, can be written as:

$$\Pr(Y^{a=1})-\Pr(Y^{a=0}) \tag{2}$$

Though it looks fancy, the superscript just indicates that we are talking about potential outcomes. $\Pr(Y^{a=1})$ is asking a hypothetical question: what neonatal mortality would be observed if all hospitals used Program A? Similarly, $\Pr(Y^{a=0})$ is asking: what neonatal mortality would be observed if all hospitals did not use Program A? Notice there are no vertical bars in the expression above. We are really comparing the neonatal mortality in all hospitals in two different scenarios. As said earlier, this is a nice definition of a causal effect but how do we estimate it with observed data?

Let us state the problem like this. We can only estimate (1) but what we really want to know is (2). Under what conditions, or assumptions, will these two things be equal?

$$\Pr(Y|A=1)-\Pr(Y|A=0)=\Pr(Y^{a=1})-\Pr(Y^{a=0}) \tag{3}$$

This is one of the major tasks of causal inference: to clarify the assumptions required to say that something observable (the left side) is equal to an unobservable causal effect (the right side). In our case, causal inference can tell us what assumptions we need to believe to be able to say that the risk difference we get when we compare the treated and untreated, is equal to the risk difference we would find if we had treated everyone and also not treated everyone. In the next section, we will go step by step to see how each causal assumption is one step on the path from the left side to the right side of (3).

**What is the causal effect of Program A on neonatal mortality?**

Now that we have clarified how to think about a causal question, we will set about answering it with data. In our dataset, half of the hospitals use Program A while the other half do not. We have measured the average neonatal mortality both in hospitals that used Program A, 2.3/1,000, and those that did not, 2.6/1,000. These values can be found in Table 1.

The half of the Table 1 that is unshaded represents the real, observable world. This is where our datasets and models estimates live. For our example, we find the observed values 2.3/1,000 and 2.6/1,000. We can take the difference of these values to get the observed difference in neonatal mortality, -0.3. This is the left side of equation (3).

The shaded half of the table contains potential outcomes that we cannot observe. Our goal is to fill in the bottom two cells of the shaded section, $\Pr(Y^{a=0})$ and $\Pr(Y^{a=1})$, which will give us the right side of equation (3), our causal risk difference. To obtain those values we will have to fill in the darker shaded four cells in the middle of the table, two at a time, using causal assumptions.

Many cells of Table 1 contain both a mathematical expression and text that describes what that expression means in words. A quick word of warning that the language can get a little tricky when talking about potential outcomes. Parsing these sentences slowly may help.

Table 1—An outline of the basic setup of our example. The white half of the table is data from the observed world and the grey half is data from the potential outcome world. Each cell has the require value written in words and in statistical notation. The two values in the bottom row are those required to estimate a causal effect.

| Observed world | | Potential outcome world | |
|---|---|---|---|
| Exposure (A) | Outcome (Y) | $Y^{a=0}$ | $Y^{a=1}$ |
| 0 | $\Pr(Y \mid A=0) = 2.6/1{,}000$<br>The observed neonatal mortality in hospitals that did not use Program A | $\Pr(Y^{a=0} \mid A=0)$: The neonatal mortality we would have observed among hospitals that did not use Program A had they not used Program A | $\Pr(Y^{a=1} \mid A=0)$: The neonatal mortality we would have observed among hospitals that did not use Program A had they used Program A |
| 1 | $\Pr(Y \mid A=1) = 2.3/1{,}000$<br>The observed neonatal mortality in hospitals that did use Program A | $\Pr(Y^{a=0} \mid A=1)$: The neonatal mortality we would have observed among hospitals that used Program A had they not used Program A | $\Pr(Y^{a=1} \mid A=1)$: The neonatal mortality we would have observed among hospitals that did use Program A had they used Program A |
| Total (average of the two cells above) | | $\Pr(Y^{a=0})$: The neonatal mortality we would have observed had all hospitals not used Program A | $\Pr(Y^{a=1})$: The neonatal mortality we would have observed had all hospitals used Program A |

## Causal assumptions when using confounder control

*Consistency*

We want to use our observed data, the neonatal mortality in hospitals using and not using Program A, to fill in cells in the potential outcome side of the table, starting with those cells in darker shading. Read the text in Table 1 carefully to understand what each cell means. Which two darker shaded cells seem like a clear choice to fill in first? The most intuitive cells to fill in are the top-left and bottom-right darker shaded cells. The bottom right, for example, is the neonatal mortality we would have observed if we had set the hospitals that did use Program A to use Program A. These hospitals did, in fact, use Program A, so we did not even have to set them to that value. We can take the value we observed in hospitals that used Program A and plug it in. The same reasoning applies to the top left cell. We can see more clearly what we are assuming by looking at what we are equating:

$$\begin{aligned} \Pr(Y|A = 0) = \Pr(Y^{a=0}|A = 0) \\ \Pr(Y|A = 1) = \Pr(Y^{a=1}|A = 1) \end{aligned} \quad (4)$$

Or more often you will see this written more compactly as:

$$Y = Y^{a} \text{ for all those with } A = a \quad (5)$$

In words, if someone was observed to have treatment $a$, then their observed outcome $Y$ is equal to the outcome we would have observed had we set their treatment to $a$. This is known as the consistency assumption. It may feel like no assumption is even being made here. We are plugging in a value for a potential outcome that we have observed directly. Or have we? Is it possible for $\Pr(Y|A = 1)$ and $\Pr(Y^{a=1}|A = 1)$ to differ? Yes, simply because the potential outcome we imagine is not always what happens in practice.

When we say we want to know the causal effect of Program A, we are imagining this program being implemented precisely as intended. Therefore, when we say $\Pr(Y^{a=1}|A = 1)$ we mean the proportion of treated people who would have had $Y$ in a hospital that had implemented Program A perfectly. This is not necessarily equal to $\Pr(Y|A = 1)$ if the hospitals that implemented Program A deviated from the protocol. Maybe there are aspects of Program A they failed to implement or implemented at the wrong time. Therefore, the neonatal mortality rate of 2.3/1,000 does not necessarily correspond to what we would have observed if Program A had been implemented perfectly.

The same can be true of $Y^{a=0}$. When we talk about the neonatal mortality we would have observed had a hospital not implement Program A, do we mean the neonatal mortality we would have observed if hospitals had used a specific alternative standard of care? Or that they did not use any part of Program A? Or that they used whatever standard of care they normally use? What if hospitals have different standards of care? Even for the control condition, we need to be clear about what we mean by $a = 0$.

For this reason, this is sometimes referred to as the "well-defined intervention" assumption [4–6]. But, as we have seen, even if we have a perfectly well-defined definition of Program A, this definition must match what happened in the data. Two additional concepts that relate to consistency but we do not have room to expand upon here are interference[7,8] and treatment variation irrelevance[9]. See the following references for more information on consistency [1,9,10].

One last note about consistency, it takes us one step from the left of equation (3) toward the right. We can write in this intermediate step:

$$\Pr(Y|A = 1) - \Pr(Y|A = 0) = \Pr(Y^{a=1}|A = 1) - \Pr(Y^{a=0}|A = 0) = \Pr(Y^{a=1}) - \Pr(Y^{a=0}) \quad (6)$$

Let us proceed by assuming this is satisfied and therefore, we will plug in the corresponding values into Table 2. Next, we will see how to fill in the remaining cells in the darker shaded region of the table.

Table 2—The consistency assumption is used to fill in the top-left and bottom-right dark grey cells.

| Observed world | | Potential outcome world | |
|---|---|---|---|
| Exposure (A) | Outcome (Y) | $Y^{a=0}$ | $Y^{a=1}$ |
| 0 | $\Pr(Y \mid A=0) = 2.6$ | $\Pr(Y^{a=0} \mid A=0) = 2.6$ | $\Pr(Y^{a=1} \mid A=0)$ |
| 1 | $\Pr(Y \mid A=1) = 2.3$ | $\Pr(Y^{a=0} \mid A=1)$ | $\Pr(Y^{a=1} \mid A=1) = 2.3$ |
| Total | | $\Pr(Y^{a=0})$ | $\Pr(Y^{a=1})$ |

*Exchangeability*

Filling in the remaining cells is bit more challenging. What value could we possibly fill in, for example, for the bottom left darker shaded cell which is the neonatal mortality we would have observed among hospitals that used Program A had we set them to not use Program A? After all, this value is unobservable by definition. Take a moment to think about what number you could fill in there and what you would need to believe to do so. Hint: If we told you that hospitals were randomized to receive or not receive Program A, would that help fill in the remaining empty cells?

Randomization should make it so that hospitals that received Program A are the same, on average, as hospitals that did not receive Program A. Therefore, if after randomization we decided to give Program A to both arms of our trial, we would expect the same neonatal mortality in both groups. We can write this as:

$$\Pr(Y^{a=1} \mid A=0) = \Pr(Y^{a=1} \mid A=1) \tag{7}$$
$$\Pr(Y^{a=0} \mid A=0) = \Pr(Y^{a=0} \mid A=1)$$

with the bottom equality corresponding to the potential outcome under control ($a = 0$). When this is true, we call the two groups exchangeable [11]. You will often see exchangeability written as:

$$Y^a \perp\!\!\!\perp A \tag{8}$$

which, in words, can be read to say that the observed treatment status A, is not correlated or associated with the potential outcomes $\Pr(Y^{a=1})$ and $\Pr(Y^{a=0})$.

If we believe the exchangeability assumption, we can use the equations in (7) to take the 2.6/1,000 in the top-left of the darker shaded cells and copy it into the cell below it (Table 3). Similarly we can take the 2.3/1,000 and copy it into the cell above it. Once we have done this, we can fill in the bottom row because the average of two equal numbers, e.g. 2.6 and 2.6, is simply that same number, 2.6.

Once we fill in the bottom row, we can see that (7) also lets us say $\Pr(Y^{a=1}|A=1)=\Pr(Y^{a=1})$ and $\Pr(Y^{a=0}|A=0)=\Pr(Y^{a=0})$. This is the last equality in equation (6) completing the last step from our observed risk difference to the causal risk difference. We made it!

You might wonder what can be done if the hospitals were not randomized and there is confounding present. It may be that tertiary hospitals, which tend to have better outcomes, were also much more likely to take up Program A than non-tertiary hospitals. As a result, we do not believe that the outcome in the hospitals without Program A is what would have happened in the hospitals with Program A had they not used Program A. In other words, exchangeability is not satisfied. In this case, we must rely on conditional exchangeability instead. Conditional exchangeability simply assumes that exchangeability is met once we have conditioned on confounders. Another way to think of conditional exchangeability is that it assumes exchangeability not overall, but within strata of confounders. If we are worried about confounding by whether a hospital is tertiary or not, we could estimate our causal effect separately in tertiary and non-tertiary hospitals and then pool the estimates. We would then need to assume exchangeability conditional on tertiary hospital status. If we are adjusting for more potential confounders, we are simply assuming exchangeability conditional on more and more variables.

You may have noticed that talking about exchangeability and confounding go hand in hand. The conditional exchangeability assumption is the equivalent to saying that, once you have adjusted for potential confounders in your analysis (and potentially also addressed selection bias in some way) that you believe there is no remaining confounding or selection bias.

Table 3—By assuming exchangeability, we can fill in values for the bottom-left and top-right dark grey cells. Once the dark grey cells are filled in, we can take the average of the column to obtain the values in the bottom row. We can then contrast the values in the bottom row to get the causal effect.

| Observed exposure (A) | Observed outcome (Y) | $Y^{a=0}$ | $Y^{a=1}$ |
|---|---|---|---|
| 0 | $\Pr(Y|A=0)=2.6$ | $\Pr(Y^{a=0}|A=0)=2.6$ | $\Pr(Y^{a=1}|A=0)=2.3$ |
| 1 | $\Pr(Y|A=1)=2.3$ | $\Pr(Y^{a=0}|A=1)=2.6$ | $\Pr(Y^{a=1}|A=1)=2.3$ |
| Total | | $\Pr(Y^{a=0})=2.6$ | $\Pr(Y^{a=1})=2.3$ |

*No measurement error*

Even if consistency and exchangeability are satisfied for the true underlying variables, the estimate of the causal effect can still be biased if those variables have been mismeasured [12]. We will add an asterisk to variables to indicate they have been mismeasured, e.g., $A^*$ and $Y^*$. In Table 4, 10 hospitals that used Program A were incorrectly recorded as not using it and 10 hospitals that did not use Program A were incorrectly recorded as using it. This means that both $A^*=1$ and $A^*=0$ are a mix of hospitals both using and not using Program A resulting in an observed risk difference of 2.36-2.54 = -0.18. If we add measurement error in the outcome, say that 5% of neonatal mortalities are incorrectly classified as stillbirth in Program A hospitals and 10% in non-Program A hospitals, the observed risk difference becomes 2.22-2.31 = -0.09. If we are unaware that these variables were mismeasured, we would think the causal effect was only -0.09 instead of the true -0.3.

Additionally, measurement error in covariates can also lead to bias. For example, if average income of patients was a confounder but it was measured with much error (because people tend to misreport their income), the mismeasured version of income will do a poorer job of controlling that confounding.

Table 4—Variables with an asterisk are mismeasured. 10 hospitals were incorrectly considered to be using Program A even though they were not (row 2) and another 10 were incorrectly considered to not be using Program A even though they were (row 5). The fourth column shows the true neonatal mortality that would be found among A*=1 and A*=0 hospitals. The fifth column adds additional measurement error where 5% of neonatal mortalities are misclassified as stillbirth in Program A hospitals and 10% of neonatal mortalities are misclassified as stillbirth in non-Program A hospitals.

| Recorded exposure ($A^*$) | True exposure (A) | Hospitals | True neonatal mortality (Y) | Recorded neonatal mortality ($Y^*$) |
|---|---|---|---|---|
| 1 | 1 | 40 | 2.3 | $2.3 \times 0.95 = 2.19$ |
| 1 | 0 | 10 | 2.6 | $2.6 \times 0.90 = 2.34$ |
| All with $A^*$=1 | | 50 | $\Pr(Y \mid A^* = 1) = 2.36$ | $\Pr(Y^* \mid A^* = 1) = 2.22$ |
| | | | | |
| 0 | 0 | 40 | 2.6 | $2.6 \times 0.90 = 2.34$ |
| 0 | 1 | 10 | 2.3 | $2.3 \times 0.95 = 2.19$ |
| All with $A^*$=0 | | 50 | $\Pr(Y \mid A^* = 0) = 2.54$ | $\Pr(Y^* \mid A^* = 0) = 2.31$ |

*Positivity*

There is a way our answer may be biased even if consistency and conditional exchangeability are satisfied and we have measured all variables without error. To illustrate this, let us pick up from our analysis stratified on whether a hospital is tertiary. Imagine that we find that no non-tertiary hospitals used Program A (Table 5). If we now try to estimate the effect in both tertiary and non-tertiary hospitals in order to pool the estimates, we would find no effect of Program A (a risk difference of 0/1,000) in tertiary hospitals but we would also find that we could not obtain an estimate among non-tertiary hospitals. This lack of hospitals using Program A among non-tertiary hospitals is a violation of the positivity assumption [13].

The positivity assumption assumes that we have both treated and untreated units, hospitals in this case, when we look in strata of our confounders. This is often written as:

$$0 < \Pr(A = 1 \mid L = l) < 1 \text{ for all observed values of } l \tag{9}$$

In words, this says that the probability of being treated is neither 0 nor 1 in all levels of confounders which here are represent by L. If this probability of being treated is 0 or 1 in any stratum of L that means that either no one is treated or everyone is treated. In our example, the probability of using Program A in non-tertiary hospitals is 0, $\Pr(A = 1|L = 0) = 0$. Therefore, positivity is violated. Why does this matter? Imagine you had not stratified your data as in Table 4 but, instead, put your hospital data into a regression model adjusting for tertiary hospital. The regression would return an effect of 0/1,000 because it would simply estimate the effect among tertiary hospitals. You would come away thinking that the program had no effect anywhere when this number only applies to tertiary hospital. If the effect of Program A is the same in non-tertiary hospitals as in tertiary hospitals, then we would be lucky because 0/1,000 would also correspond to the causal effect in all hospitals. But if the effect of Program A is different in non-tertiary hospitals, then 0/1,000 is not the true effect among all hospitals. Our effect estimate will be biased.

Positivity violations come in two types. The first occurs when it is not possible for some to be treated (or untreated). In our example this would be the case if only tertiary hospitals had the infrastructure to implement the program because it would be impossible for non-tertiary hospitals to use Program A. The second type occurs when no treated or untreated units are observed by chance. In our example this would be the case if very few non-tertiary hospitals were included in the dataset and, by chance, all of them did not use Program A. Each type has different consequences for how the results can be interpreted or how it might be addressed[13].

Some positive news about the positivity assumption, it can be checked either by stratifying as above or, when there are many confounders, by using propensity scores. You can read more about the positivity assumption here [13] including how one might address positivity violations [14].

Table 5—Repeating the above steps within strata of whether the hospital is tertiary. In this example, positivity is not satisfied because there are no non-tertiary hospitals that used Program A.

Tertiary hospitals (L=1)

| Observed exposure (A) | Observed outcome (Y) | $Y^{a=0}$ | $Y^{a=1}$ |
|---|---|---|---|
| 0 | $\Pr(Y\|A = 0, L = 1) = 2.3$ | $\Pr(Y^{a=0}\|A = 0, L = 1) = 2.3$ | $\Pr(Y^{a=1}\|A = 0, L = 1) = 2.3$ |
| 1 | $\Pr(Y\|A = 1, L = 1) = 2.3$ | $\Pr(Y^{a=0}\|A = 1, L = 1) = 2.3$ | $\Pr(Y^{a=1}\|A = 1, L = 1) = 2.3$ |
| Total | | $\Pr(Y^{a=0}\|L = 1)$=2.3 | $\Pr(Y^{a=1}\|L = 1) = 2.3$ |

Non-tertiary hospitals (L=0)

| Observed exposure (A) | Observed outcome (Y) | $Y^{a=0}$ | $Y^{a=1}$ |
|---|---|---|---|
| 0 | $\Pr(Y\|A = 0, L = 0) = 2.9$ | $\Pr(Y^{a=0}\|A = 0, L = 0) = 2.9$ | $\Pr(Y^{a=1}\|A = 0, L = 0) = NA$ |
| 1 | $\Pr(Y\|A = 1, L = 0) = NA$ | $\Pr(Y^{a=0}\|A = 1, L = 0) = 2.9$ | $\Pr(Y^{a=1}\|A = 1, L = 0) = NA$ |
| Total | | $\Pr(Y^{a=0}\|L = 0)$=2.9 | $\Pr(Y^{a=1}\|L = 0) = NA$ |

## Concluding thoughts

This has been a quick and basic introduction to the four causal assumptions required when estimating causal effects by adjusting for confounding. Of course, an actual causal analysis would not explicitly go through the steps as we have done. The purpose of going through these steps to give some intuition for what the causal assumptions are and how, when they are satisfied, an observed association could be interpreted as causal. Nonetheless, the assumptions presented here play a similar role regardless of whether one is using the Mantel-Haenszel method or g-computation. In a study with causal goals these assumptions should not only be mentioned by name, but in the language of the study at hand. For example, instead of simply saying, "this study assumes conditional exchangeability," one could be more specific by saying, "this study assumes that adjusting for hospital tertiary status is sufficient to control all confounding and selection bias between the use of Program A and neonatal mortality. We also assume no selection bias occurred in the selection of hospitals." It is also important that a causal effect can only be inferred if you have strong arguments for why the reader should believe the assumptions. These arguments can come in the form of expert opinion about the adjustment variables (possibly using directed acyclic graphs[15]) or quantitatively in the form of bias analyses [16], negative controls [17] or triangulation[18].

It should be noted that having well-specified models is often included in the list of assumptions needed to estimate causal effects, i.e., that the variables in the model are being correctly modelled. For example, if adjusting for maternal age and the relationship between maternal age and neonatal mortality is not linear but it was modelled linearly in an outcome model, that model would be misspecified and would likely produce a biased estimate. Another example would be that the relationship of the hospital type and neonatal mortality depends on maternal age, any outcome model that is missing that statistical interaction would be misspecified and, therefore, likely produce a biased estimate. Though it is clearly a necessary assumption, it is often seen as a statistical assumption and not a causal assumption.

Another simplification we have made is that we have assumed no selection bias. As mentioned, selection bias is part of the exchangeability assumption and though it can be addressed through confounder control, there are other ways it can be addressed that were not mentioned here such as inverse probability of censoring weighting[19].

One interesting point is that these assumptions are the same required in randomized controlled trials. The advantage of randomized controlled trials is that it is much easier to convince the reader that these assumptions are satisfied. As mentioned, exchangeability is expected to be satisfied on average due to randomization. Consistency is also expected to be satisfied when the protocol of the randomized trial is well-defined and the protocol is closely adhered to. Similarly, positivity is satisfied in a typical randomized trial because, even if we condition on covariates, there is no reason to expect only treated or untreated people within strata of confounders.

The assumption set outlined above is very common in the health sciences to infer causation from observed data though it is not the only assumption set to infer causation. There are many other assumption sets which, if satisfied, can also infer causation such as instrumental variable analysis, regression discontinuity and difference-in-differences.

We hope that this article provides deeper intuition for why the causal assumptions of consistency, exchangeability, no measurement error and positivity are able to accomplish the impressive task of inferring causation from observable data. Additionally, we hope that this also leads to their inclusion in studies with causal goals.

**Contributors and sources:** JAL conceived and wrote the article. JAL is Assistant professor of epidemiology and head of the Causal Inference Group at Erasmus MC. JAL is guarantor of this article.

**Competing interests:** JAL declares no competing interests.

**Funding:** The article has no relevant funding.

**Patient and public involvement:** There was no patient or public involvement in this manuscript.